\documentclass{IEEE_lsens}
\usepackage{amsmath,amsfonts}
\usepackage{algorithmic}
\usepackage{textcomp}
\usepackage{multirow}
\usepackage[table,xcdraw]{xcolor}
\usepackage{graphicx}
\usepackage{array}
\usepackage{caption}
\usepackage{subcaption}

\usepackage[noadjust]{cite}

\ifCLASSINFOpdf
\else
\fi
\usepackage[T1]{fontenc} 
\usepackage{amsmath}
\usepackage[cmintegrals]{newtxmath}
\usepackage{bm}

\usepackage{array}
\usepackage{url}
\ifCLASSINFOpdf
\else
\fi
\providecommand{\hypersetup}[1]{\relax}

\hypersetup{pdftitle={Bare Demo of IEEE\_lsens.cls for IEEE Sensors Letters},
pdfsubject={Typesetting},
pdfauthor={Michael D. Shell},
pdfkeywords={Class, IEEE, IEEE\_lsens, IEEE Sensors Letters, LaTeX, Typesetting, TeX}}
\begin{document}

%
\title{Insights into Age-Related Functional Brain Changes during Audiovisual Integration Tasks: A Comprehensive EEG Source-Based Analysis}

%
\author{\IEEEauthorblockN{Prerna Singh\IEEEauthorrefmark{1}, Ayush Tripathi\IEEEauthorrefmark{2}, 
 Lalan Kumar\IEEEauthorrefmark{1,2,3} and Tapan Kumar Gandhi\IEEEauthorrefmark{1,2}}
\IEEEauthorblockA{\IEEEauthorrefmark{1} Bharti School of Telecommunication Technology and Management, Indian Institute of Technology Delhi, Delhi 110016, India\\
\IEEEauthorrefmark{2}Department of Electrical Engineering , Indian Institute of Technology Delhi, Delhi 110016, India\\
\IEEEauthorrefmark{3}Yardi School of Artificial Intelligence, Indian Institute of Technology Delhi, Delhi 110016, India\\
\\}%
}
%
%
%

\IEEEtitleabstractindextext{%
\begin{abstract}
The seamless integration of visual and auditory information is a fundamental aspect of human cognition.
Although age-related functional changes in Audio-Visual Integration (AVI) have been extensively explored in the past,
thorough studies across various age groups remain insufficient. Previous studies have provided valuable insights into
age-related AVI using EEG-based sensor data. However, these studies have been limited in their ability to capture
spatial information related to brain source activation and their connectivity. To address these gaps, our study conducted
a comprehensive audio-visual integration task with a specific focus on assessing the aging effects in various age groups,
particularly middle-aged individuals. We presented visual, auditory, and audio-visual stimuli and recorded EEG data from
Young (18-25 years), Transition (26-33 years), and Middle (34-42 years) age cohort healthy participants. We aimed to
understand how aging affects brain activation and functional connectivity among hubs during audio-visual tasks. Our
findings revealed delayed brain activation in middle-aged individuals, especially for bimodal stimuli. The superior temporal
cortex and superior frontal gyrus showed significant changes in neuronal activation with aging. Lower frequency bands
(theta and alpha) showed substantial changes with increasing age during AVI. Our findings also revealed that the AVI-
associated brain regions can be clustered into five different brain networks using the k-means algorithm. Additionally, we
observed increased functional connectivity in middle age, particularly in the frontal, temporal, and occipital regions. These
results highlight the compensatory neural mechanisms involved in aging during cognitive tasks.
\end{abstract}

}

\maketitle

\section{Introduction}
In our daily lives, we encounter a multitude of stimuli from different sensory modalities, such as auditory, vision, smell, and touch.  Our remarkable brain has the capacity to efficiently process and integrate pertinent information from these different sources, which is known as multisensory integration. This allows us to perceive and understand the external world effectively amidst the dynamic and complex information surrounding us. When communicating with others, we integrate visual and auditory information to understand speech content. This integration process is known as audio-visual integration \cite{laurienti2006enhanced,stein2012new}. Audio-visual stimuli elicit faster and more accurate responses compared to unimodal stimuli, highlighting the effectiveness of audiovisual integration \cite{teder2005effects,yang2015effects}.

\par With advancing age, there is a noticeable decline in both sensory systems and cognitive functions \cite{backman2006correlative,mozolic2012multisensory}. The process of audiovisual integration serves as a bridge, connecting sensory and cognitive processing, thus mitigating the impact of age-related declines in both domains \cite{mcgovern2014sound}. Age-related declines in cognitive function contribute to increased auditory threshold and decreased visual acuity in older adults. Interestingly, studies on audio-visual integration have shown that older adults exhibit an enhanced audio-visual integration effect compared to younger adults in tasks involving auditory/visual discrimination \cite{peiffer2007age,zou2017aging}, like sound-induced flash illusion tasks \cite{deloss2013multisensory}  and speech perception task \cite{sekiyama2014enhanced}. Emerging evidence from these studies suggests that audiovisual integration (AVI) may serve as a compensatory mechanism to counteract functional decline associated with aging. However, contrary findings have also been extensively documented in studies employing tasks such as the auditory/visual discrimination tasks \cite{ren2017audiovisual,wu2012age}, and the sentence discrimination task \cite{tye2010aging}.

\par Additionally, the temporal aspect of audiovisual integration (AVI) plays a vital role in determining the occurrence of integration. Past researchers have found that the window for binding is extended for complex stimuli compared to simpler audiovisual stimuli \cite{stevenson2013multisensory}. A compelling finding from a previous study \cite{ren2018comparison}  revealed that a powerful multi-sensory integration effect is observed when the temporal gap between auditory and visual stimuli is less than 100 milliseconds. Hence, variations in experimental materials have been suggested as a primary factor contributing to the contrasting results observed in these studies. Furthermore, the location of the stimulus, whether presented peripherally \cite{mahoney2011multisensory, wu2012age} or centrally \cite{diaconescu2013visual, zou2017aging}, has varied across studies, leading to differing outcomes. Given the significant age-related decline in peripheral perceptual processing, the specific presentation location of the stimuli has also contributed to the conflicting findings.

\par Subsequent ERP studies on AVI elucidate a compelling insight that older adults displayed a heightened neural response to audiovisual stimuli, particularly in the medial prefrontal and inferior parietal regions \cite{ren2018comparison}. These findings strongly supported the notion that the amplified audiovisual integration observed in older adults serves as a compensatory mechanism, counteracting deficiencies in unimodal sensory processing \cite{mozolic2012multisensory}.

\par Past Studies have found that neural oscillatory responses in various frequency bands, such as theta, alpha, beta, and gamma, play a role in sensory processing \cite{donner2011framework}. Theta and alpha bands, particularly in fronto-centro-parietal sites, are involved in cognitive control, short-term memory, sensory information maintenance, and suppression of distractions \cite{sakowitz2005spatio, kawasaki2010dynamic}. Studies on aging based on EEG have revealed a decline in alpha power, indicating age-related differences in audiovisual integration within the low-frequency bands (theta and alpha).

\begin{figure*}[t]
\centering
\begin{subfigure}[h]{0.5\linewidth}
\includegraphics[width =\linewidth]{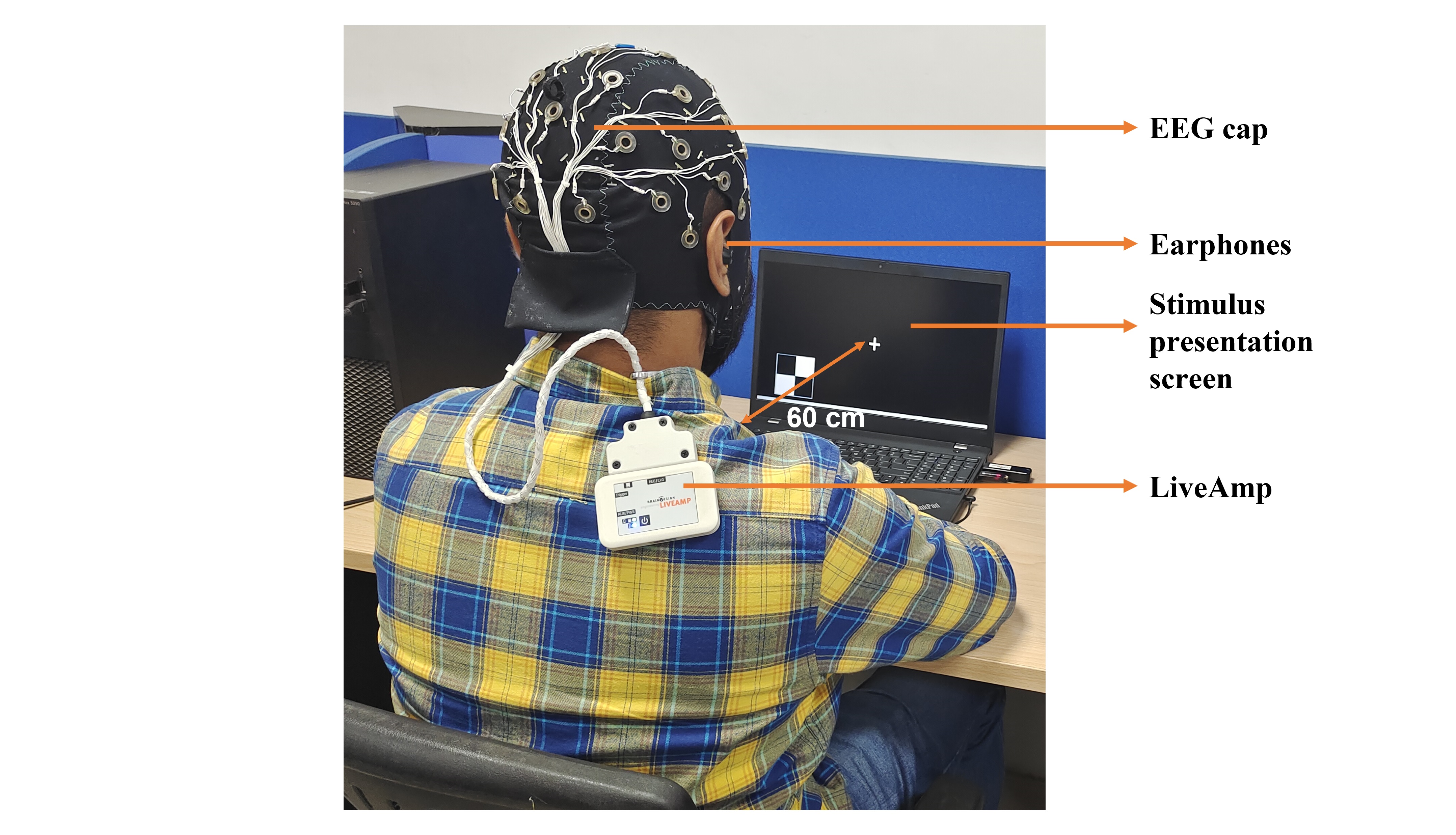}
\caption{Depiction of the experimental setup.}
\label{figure: Experimental setup}
\end{subfigure}
\begin{subfigure}[h]{0.49\linewidth}
\includegraphics[width =\linewidth]{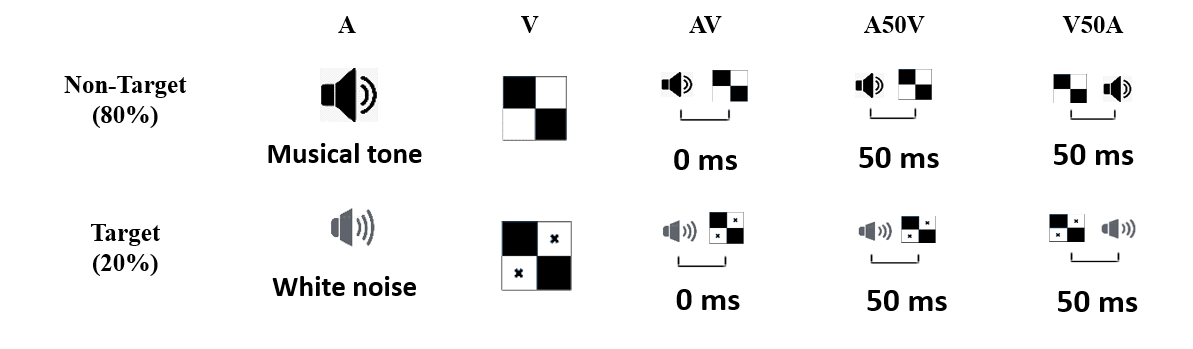}
\caption{Description of different stimuli involved during data collection. } 
\label{figure: Stimuli description}
\end{subfigure}
\begin{subfigure}[h]{0.95\linewidth}
\includegraphics[width =\linewidth]{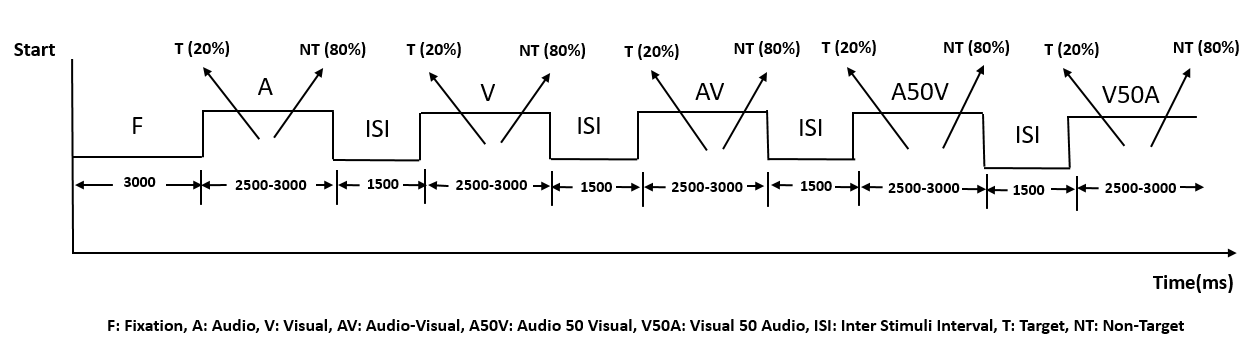}
\caption{Experimental timeline of one session.} 
\label{figure: Experimental timeline}
\end{subfigure}
\caption{Schematic of the experimental design.}
\label{figure 1}
\end{figure*}

\par While several models successfully explain behavioral indices of audiovisual integration, the precise neural mechanisms underlying efficient integration in the brain remain unclear. In a meta-analytic study, common brain activity patterns were identified across diverse audiovisual studies \cite{gao2023audiovisual}. They discovered that unisensory signals are processed independently in sensory cortices, while integration occurs in later association areas like the superior temporal cortex. However, accumulating evidence suggests that integration might occur at sensory-perceptual and sub-cortical levels prior to the involvement of higher association cortices \cite{gao2023audiovisual}. Some researchers have used PET scans to identify that the right insula is most strongly involved in audiovisual synchrony-asynchrony detection \cite{bushara2001neural}.
So, it is clear that audiovisual integration involves various brain regions, including occipital, parietal, temporal, and frontal areas \cite{shams2005early}. Studies using fMRI and EEG have observed integration effects even in sensory-specific regions like the primary visual cortex \cite{macaluso2006multisensory,chan2017spatio,ren2017audiovisual}. Anatomical connections between occipital and superior temporal regions highlight their crucial roles in audiovisual integration \cite{falchier2002anatomical}. Functional connectivity, characterized by temporal correlations or synchronization of physiological signals, provides insights into the coordination among widely distributed brain regions \cite{bhattacharya2018functional,wang2018mulan}.

\par Recent studies emphasize the role of functional connectivity in cognitive functioning \cite{wen2012causal}. In multi-sensory processing, these connections are crucial for integrating information between sub-cortical structures and cortical areas \cite{beer2011diffusion,van2014subcortical}. Studies have explored how functional networks influence audiovisual integration. However, the age-related differences in functional connectivity during audiovisual integration remain unknown.
Past graph theoretical analysis of EEG and MEG data has demonstrated that aging leads to alterations in functional connectivity and network efficiencies \cite{smit2008heritability, vecchio2014human}. Older adults exhibit increased functional connectivity and higher brain network efficiencies during synchronous audiovisual integration in the beta band \cite{wang2017beta}. However, the impact of aging on functional connectivity during audiovisual integration tasks with temporally asynchronous stimuli requires further investigation.
\par Prior research has predominantly focused on examining functional connectivity in older age groups in the sensor domain, leaving a significant gap in understanding the changes in functional connectivity during audiovisual integration tasks in the middle-aged population. Exploring these changes in middle-aged individuals in the source domain is crucial as it can give spatial information that can aid in the early detection of disorders like mild cognitive impairment (MCI) and contribute to the early identification of Alzheimer's disease (AD), which typically manifests in later stages of life. Detecting MCI in middle age is particularly important as the patho-physiological process begins years before the onset of dementia. However, investigations targeting middle-aged groups are limited, and there is a need to shift the focus towards MCI identification in this population to identify early bio-markers of cognitive decline \cite{kremen2014early,sperling2011toward}. Additionally, the application of source modeling is crucial for a more precise interpretation of variations between the young and middle-aged groups. Sensor-level data may lack spatial precision, but source modeling provides valuable information about the timing and precise brain regions involved \cite{michel2004eeg,baillet2017magnetoencephalography}. This approach aids in resolving any uncertainties inherent in sensor-level analysis. 

\begin{figure*}[t]
\centering
\includegraphics[width =0.5\linewidth,keepaspectratio]{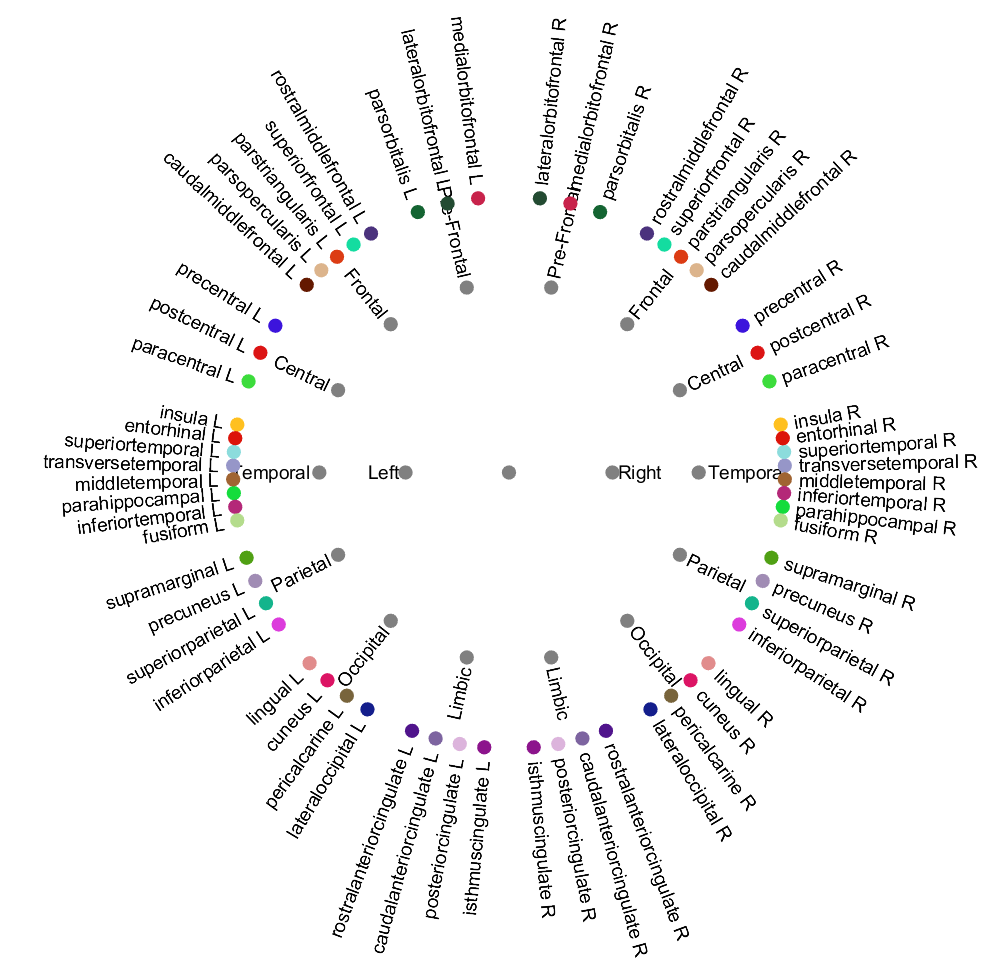}
\label{figure: figure 2}
\caption{Brain lobes divided into 62 scouts used for connectivity analysis.}
\end{figure*}

\par Motivated by this, the present study investigates brain source activation and functional connectivity in different age cohorts during audiovisual integration tasks. Participants are divided into three age groups: young (18-25 years), transition from young to middle age (26-33 years), and middle age (34-42 years). They perform auditory and visual discrimination tasks using unimodal and bimodal stimuli. EEG signals are collected from different brain regions to analyze the source-based functional connectivity network. The study examines changes in brain sources related to age and task. Features extracted from these networks are combined, and Machine Learning models are used to classify participants into distinct age groups.
\par The rest of the paper is structured into the following sections. Section II describes the materials and methods. Section III presents the results. Section IV discusses the findings, and Section V concludes the paper.

\maketitle

\section{Materials and Methods}

\subsection{Participants}
In this study, a total of fifteen  healthy subjects participated, and each subject completed five different sessions as part of the study. These participants were divided into three age groups. The first group consisted of young subjects aged 18-25 years (mean age ± SD: 21 ± 1.41 years, n=5). The second group included individuals in the transition from young to middle age, aged 26-33 years (mean age ± SD: 27.4 ± 2.57 years, n=5). The third group consisted of middle-aged subjects aged 34-42 years (mean age ± SD: 37 ± 2.8 years, n=5). The study included participants who were students/residents of IIT Delhi and met specific criteria. They had normal or corrected-to-normal vision. None of the participants had color blindness and hearing threshold issues. Importantly, all participants were unaware of the purpose of the experiment, ensuring unbiased and objective responses. All participants in the study had normal cognitive functioning, as indicated by their Mini-Mental State Examination (MMSE) scores  \cite{bravo1997age}, which were above 24. They also had no known history of cognitive impairment. Prior to participating, all individuals provided written informed consent, and the study protocol was approved by the Institute Ethics Committee of IIT Delhi (Reference $\#2021/P052$).

\subsection{Stimuli and Task}
During the experiments, two types of auditory and visual stimuli were utilized: target and non-target stimuli. The non-target visual stimulus used in the experiment was a black-and-white checkered box. On the other hand, the target visual stimulus featured a black-and-white checkered box with two cross markings inside it. The dimensions of the target visual stimulus were 52 mm × 52 mm, with a visual angle of 5\textdegree. The non-target auditory stimulus was a musical sound, while the target auditory stimulus was white noise.
The visual stimuli (V) were presented on a black screen of a 15.6-inch laptop positioned 60 cm in front of the participants' eyes. They were shown randomly on the screen for 150 ms in either the lower left or lower right quadrant. The auditory stimuli (A) were delivered through earphones at a sound pressure level of 60 dB to either the left or right ear for a duration of 150 ms. Figure 1 offers the schematic of the experimental design. A visual representation of the experimental setup can be seen in Figure \ref{figure: Experimental setup}.

The experiment included three types of stimuli: unimodal audio (A), unimodal visual (V), and audio-visual stimuli (AV). In the audio-visual condition, the stimuli were presented in three different ways based on the stimulus onset Asynchrony (SOA). The three ways for audio-visual stimulus presentation include simultaneous audio-visual (AV) where the audio and visual stimuli were presented simultaneously, Visual lag auditory by 50 ms (V50A) where the visual stimulus appeared 50 ms after the auditory stimulus, and Auditory lag visual by 50 ms (A50V) where the auditory stimulus appeared 50 ms after the visual stimulus.
Each trial of every stimulus lasted between 150 to 250 ms, with the specific duration determined based on prior behavioral investigations \cite{yang2014elevated}. Participants were instructed to perform a discrimination task involving visual (V), auditory (A), and audio-visual (AV) stimuli, both in synchrony and asynchrony. The description of the stimuli can be seen in Figure \ref{figure: Stimuli description}.

Each participant underwent five sessions, each starting with a fixation time of 3000 ms. Following that, the screen displayed 20 stimuli for each of the five types (A, V, AV, A50V, V50A). In each condition, either on the left or right side of the screen, 80\% of the stimuli were non-targets and 20\% were targets. The inter-stimulus interval between consecutive stimuli ranged from 1300 to 1800 ms.
Participants were instructed to identify whether the targets appeared on the left or right side of the screen. They were asked to do this as quickly and accurately as possible by pressing the left arrow key for left-side targets and the right arrow key for right-side targets. The experiment timeline is illustrated in Figure 1c.

\subsection{Data Collection}
Both EEG data and behavioral data were acquired simultaneously in a dimly lit room. Stimulus presentation and collection of behavioral responses were done using PsychoPy-2022.1.3 \cite{peirce2009generating}. EEG signals were recorded via a cap (EasyCap from Brain Products) equipped with 32 scalp electrodes, following the International 10–20 System. Impedance was maintained below $20 k\Omega$. The raw signals were digitized at a sampling rate of 500 Hz using LiveAmp amplifiers (BrainProducts, Munich, Germany). All data were digitally stored for subsequent offline analysis.

\subsection{Data Analysis}
\subsubsection{EEG Data Pre-Processing}
The EEG data was pre-processed using MATLAB R2021a (MathWorks, Inc., Natick, MA, United States) with the open-source EEGLAB toolbox (Swartz Center for Computational Neuroscience, La Jolla, CA, United States) \cite{delorme2004eeglab}. The pre-processing part focused on non-target stimuli to eliminate motor response and decision-making effects. Initially, the EEG data from IO channel were excluded in order to reduce the noise related to eye movements. Later, the continuous EEG data were bandpass filtered between 0.5 and 40 Hz. Further, the data were re-referenced to average reference. After that, Independent Component Analysis (ICA) \cite{delorme2004eeglab} was used to remove artifacts including eye artifacts, frequency interference, muscle artifacts, head movement, and electrocardiographic activity \cite{jung1998analyzing,delorme2004eeglab}. The data was further divided into epochs with time points ranging from 500 ms before stimulus onset to 900 ms after stimulus start. At last, Baseline correction was applied. This produced $80$ epochs (non-target stimuli only) with $700$ time points for each stimulus type per participant. These epochs were further used for data analysis.

\subsubsection{Source Domain Analysis}
We have used Brainstorm for source domain analysis \cite{tadel2011brainstorm}. It provides comprehensive source estimation tools for in-depth analysis at both individual and group levels. EEGLAB primarily focuses on sensor-level analysis and statistical modeling, rather than bio-physiological sources.  Therefore, we integrated EEGLAB's preprocessing (including ICA artifact reduction) and sensor-level analysis with Brainstorm's source modeling using the preprocessed data.
\par Then cortical source activations were estimated later \cite{ stropahl2018source,pani2020subject}. Brainstorm employs a distributed dipoles model for fitting. In our experiment, we utilized the Standardized Low-Resolution brain Electromagnetic Tomography approach (SLORETA) by Pasqual-Marqui $(2002)$ to analyze the data \cite{pascual2002standardized}. sLORETA normalizes default current density maps using data covariance, calculated from a combination of noise and brain signal covariance. The sLORETA method employs minimum-norm imaging to estimate scalp-recorded electrical activity locations.
 \par Following EEGLAB-based pre-processing (including artifact attenuation, filtering, and epoching), EEG data was imported into Brainstorm. Source estimation was confined to the cortex volume and projected onto the Montreal Neurological Institute (MNI) ICBM$152$ brain template \cite{fonov2011unbiased} using a multi-linear registration technique within Brainstorm. The ICBM152 anatomical template was used to create the forward model \cite{rodriguez2024unveiling}. Single-trial pre-stimulus baseline intervals from -500 ms to 0 ms were employed to compute single subject noise covariance matrices and derive individual noise standard deviations at each location as described by \cite{10.1093/acprof:oso/9780195307238.001.0001}. The boundary element method (BEM), implemented in OpenMEEG \cite{gramfort2010openmeeg}, served as a head model using Brainstorm's default settings. Source estimation involved selecting the option of constrained dipole orientations, where a dipole was modeled for each vertex, oriented perpendicular to the cortical surface \cite{tadel2011brainstorm}.\par Prior to source estimation, EEG data were re-referenced to the common average, a standard pre-processing step in source analysis software. Re-referencing to the common average is done to meet the assumption of zero current flow for unbiased source strength estimates \cite{michel2004eeg}. Single-trial EEG data is averaged per participant, and source estimation is conducted on the subject average. The cortical surface is divided into regions of interest (ROIs) using the Mindboggle structural atlas \cite{klein2012101}. Source estimation was conducted for every participant in the Y, T, and M age groups. Results were averaged within each age group to assess source-level activation differences. The study also examined source activation across various frequency bands.
 \vspace{2mm}

\begin{figure*}[t]
\centering
\begin{subfigure}[h]{0.8\linewidth}
\includegraphics[width =\linewidth]{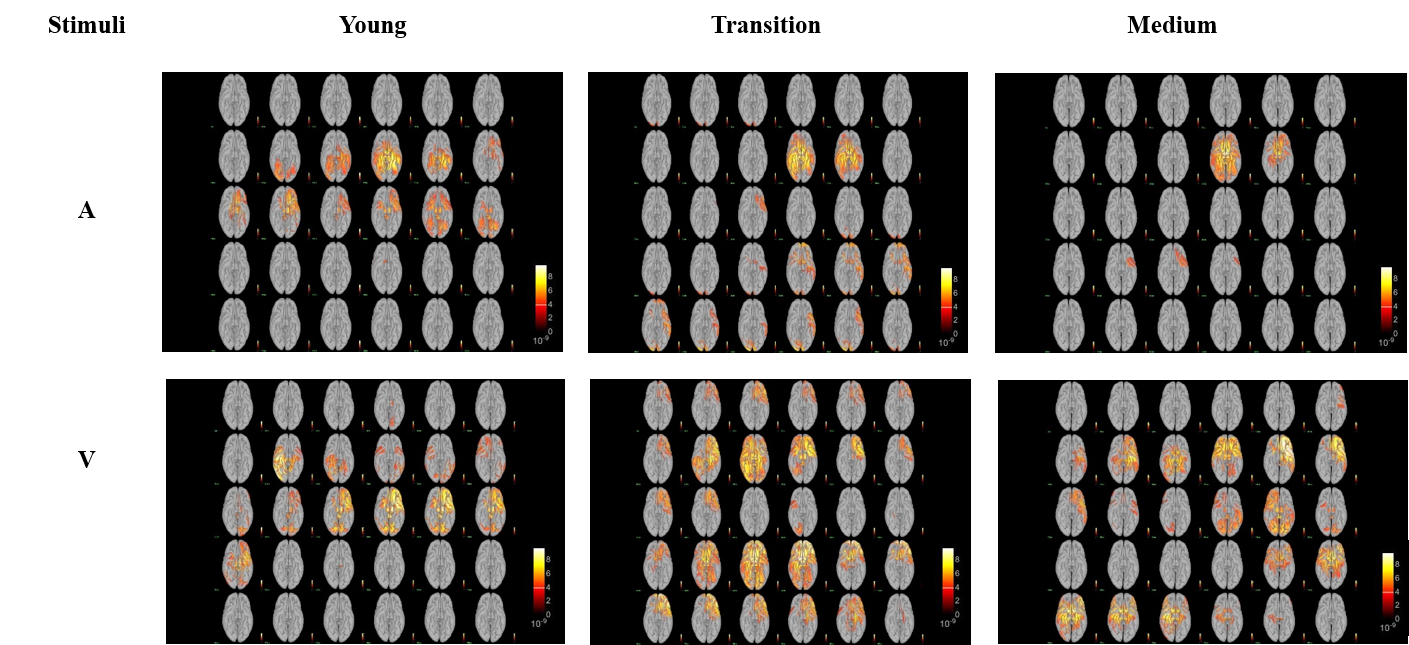}
\label{figure: figure 2a}
\caption{Brain source activation during unimodal stimuli (A and V).}
\end{subfigure}
\begin{subfigure}[h]{0.8\linewidth}
\includegraphics[width =\linewidth]{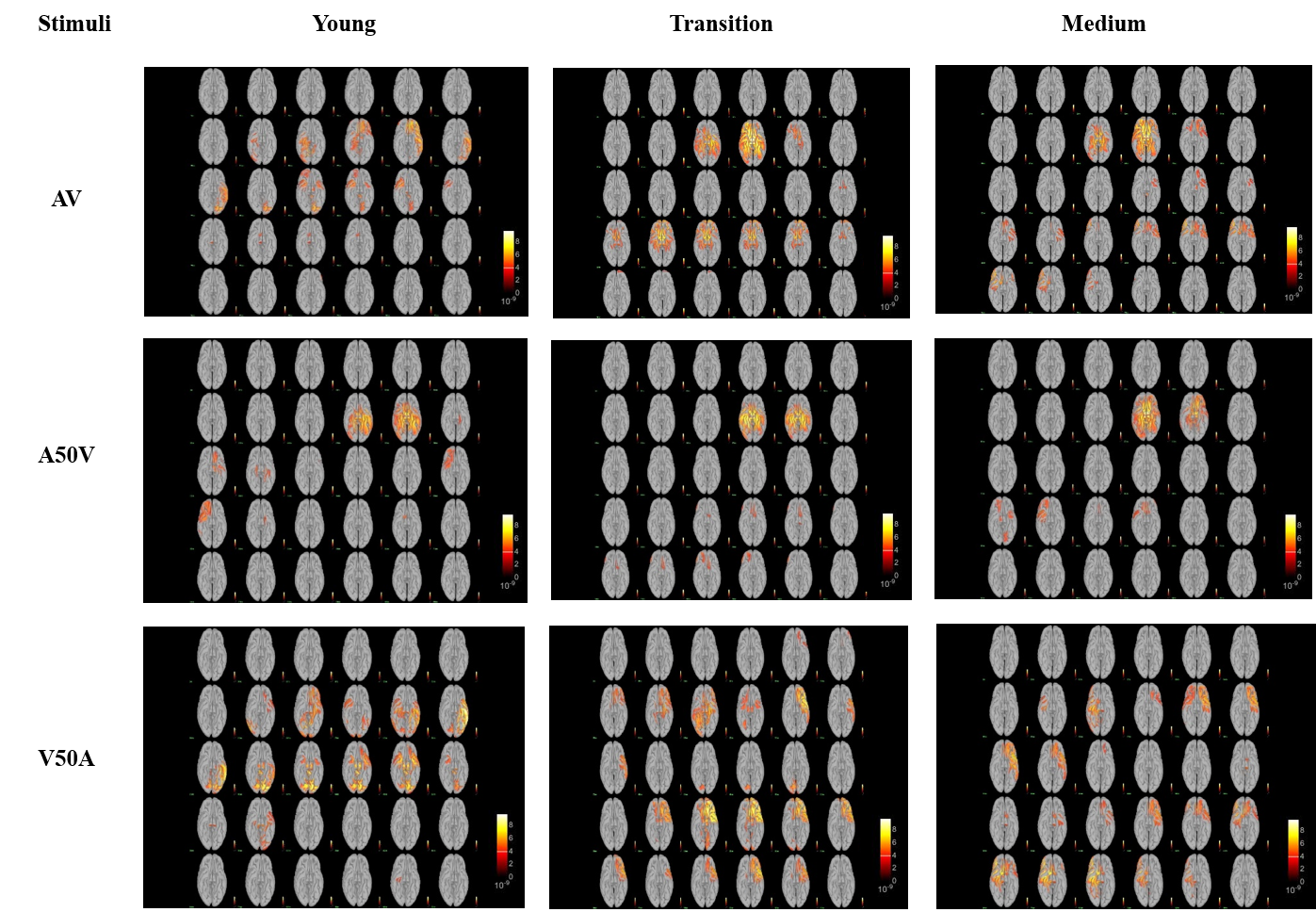}
\label{figure: Figure 2b}
\caption{Brain source activation during bimodal stimuli (AV, A50V, and A50V). } 
\end{subfigure}
\caption{Brain source activation for different time stamps using Brainstorm. Note: Each block represents brain activation from 0 to 30 time stamps covering a time window of 0 sec to 900 sec.}
\end{figure*}

\subsubsection{Extracting Source Activity Time Series}
Brainstorm offers predefined scouts (atlas-based) or manual region-of-interest (ROI) definitions. The Mindboggle structural atlas \cite{klein2012101} in Brainstorm was used to define ROIs since individual anatomies weren't available. Significant ROIs were selected for each stimulus based on scout activation time series data \cite{kouti2022emotion} and prior studies \cite{gao2023audiovisual}. Group activation averages were calculated and regions with the highest peaks in scout time series were chosen as significant audio-visual integration (AVI) scouts. These scouts were combined based on source activity to create an AVI scout for comparing age groups' AVI effects. The current approach involved manual scout definition through visual examination \cite{stropahl2018source}.  Significant scouts like the caudal middle frontal, and superior frontal, insula, superior temporal, transverse temporal, middle temporal, fusiform, parsopercularis, superior parietal were merged to create AVI scouts for analysis.

 \vspace{2mm}
\subsubsection{Time-Frequency Decomposition}
Brainstorm employed complex Morlet wavelets for Time-Frequency decomposition of brain signals. Some EEG signal aspects are challenging to assess in the time domain due to amplitude differences. Oscillations at specific frequencies carry important information, but their amplitude can be lower, making them hard to observe. Averaging in the time domain might cancel such oscillations if they lack strict phase alignment across trials. Time-frequency averaging extracts oscillation power regardless of phase shifts.

Complex Morlet wavelets are widely used in EEG analysis for time-frequency decomposition. They are sinusoidal with a Gaussian kernel, capturing local oscillatory components.

Following sLORETA-based significant scout estimation for different stimuli within each age group, Time-Frequency decomposition was performed on the significant AVI scout signals. Morlet wavelets with a mean scout function were applied to each age group's brain signals, and spectral flattening was executed. Subsequently, z-score normalization was applied to compare the results.

\subsection{Experimental Details}

\subsubsection{ Adjacency Matrix Formulation}
Using the pre-processed EEG signals, epochs were extracted for each stimulus case. All eighty epochs of every stimulus were utilized to estimate cortical sources using sLORETA 
 \cite{pascual2002standardized}. This involves creating a head model using OpenMEEG \cite{gramfort2010openmeeg}, which is then employed to derive 62 Mindboggle \cite{klein2012101} scout time series mean values for a time window of -500 ms to 900 ms. This results in $62 \times 700$ dimensional matrices, which are transposed and normalized. Each epoch is represented by a $700 \times 62$ dimensional matrix (V), with 700 representing time points and 62 representing scout numbers. These matrices form the basis for constructing adjacency matrices. The construction of a $62 \times 62$ dimensional adjacency matrix, denoted as $\hat{A}$, relies on Pearson correlation between signals from different pairs of scouts. If $\mu_{V(:,i)}$ represents the mean of the $i^{th}$  scout time series, the elements of the adjacency matrix  $\hat{A}$ are computed as:
\begin{equation}
    \hat{A}(i,j) = \frac{\sum_{l=1}^{700} (V(l,i)-\mu_{V(:,i)})(V(l,j)-\mu_{V(:,j)})}{\sqrt{\sum_{l=1}^{700} (V(l,i)-\mu_{V(:,i)})^2 \sum_{l=1}^{700}(V(l,j)-\mu_{V(:,j)})^2}}
\end{equation}

Following this, a binary adjacency matrix ($A$) is formed by utilizing the elements of $\hat{A}$ in the following manner:

\begin{equation}
A(i,j) =  
    \begin{cases}
    1 ,& \text{if } i\neq j \,\, \text{and } \hat{A}(i,j)\geq \rho_{th}\\
    0 ,& \text{otherwise } 
\end{cases}
\end{equation}

\begin{figure*}[t]
\centering
\begin{subfigure}[h]{0.49\linewidth}
\includegraphics[width =\linewidth]{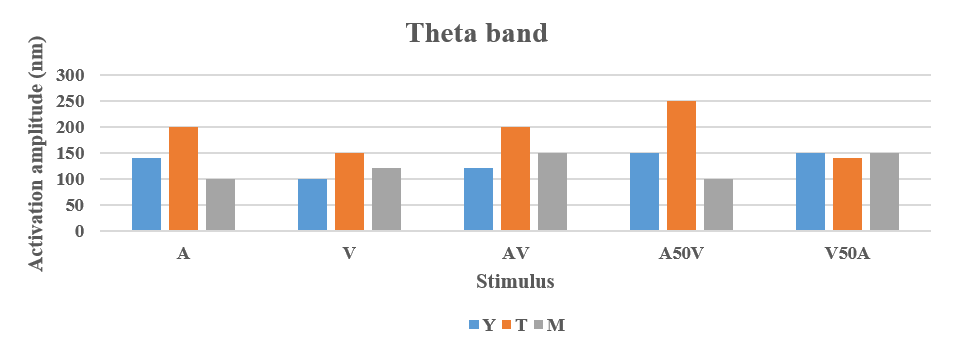}
\label{figure: figure 3a}
\caption{Activation amplitude Vs. age groups in different stimulus conditions.}
\end{subfigure}
\begin{subfigure}[h]{0.49\linewidth}
\includegraphics[width =\linewidth]{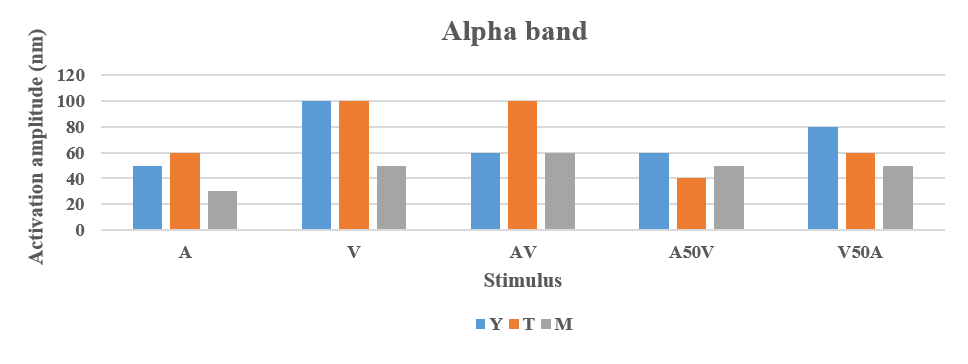}
\label{figure: Figure 3b}
\caption{Activation amplitude Vs. age groups in different stimulus conditions.} 
\end{subfigure}
\caption{Brain source activation amplitude in lower frequency bands for different age groups Y,T and M
.}
\end{figure*}

The obtained adjacency matrix will be used to construct a functional connectivity graph for a specific epoch. The threshold parameter ($\rho_{th}$) determines the correlation strength to establish an edge between two nodes. A lower $\rho_{th}$ value results in numerous edges, whereas a higher value leads to edges being formed only between nodes with a substantial signal correlation. An optimum ($\rho_{th}$) value will be used to construct a brain connectivity graph.
\vspace{2mm}

\subsubsection{Feature Extraction}

Five distinct node-level features are derived from the calculated adjacency matrix to facilitate scout-based age-group classification. This accumulates to $5 \times 62 = 310$ features for the connectivity graph of a specific epoch. The subsequent features are extracted using the \textit{NetworkX} package \cite{SciPyProceedings_11}:

\begin{enumerate}
    \item Degree Centrality: Degree centrality for a node $u$, quantifies the number of edges connected to it.

 Mathematically, it is computed as,
    \begin{equation}
        d_u = \frac{\sum_{l=1}^{N}A(u,l)}{N-1}
    \end{equation}
    where $N$ represents the total number of nodes in the graph.
    
\item Betweenness Centrality: It's used for identifying how much impact a node holds in regulating information flow within a graph. This is valuable for pinpointing nodes that connect different sections of a graph. The algorithm calculates the shortest routes between all node pairs. For a specific node $u$, it's calculated as:
    \begin{equation}
        b_u = \sum_{a,b\in U}\frac{\sigma(a,b|u)}{\sigma(a,b)}
    \end{equation}

In this equation, $U$ represents the set of nodes, $\sigma(a,b)$ indicates the count of shortest paths connecting nodes $a$ and $b$, and $\sigma(a,b|u)$ represents these shortest paths that go through node $u$. It's worth noting that $\sigma(a,b)$ is 1 when $a$ and $b$ are the same, and $\sigma(a,b|u)$ is 0 when $u\in a,b$.

\item Eigenvector Centrality: It is a measure of a node's significance in a graph based on the importance of its connected neighbors. Nodes connected to more important nodes receive higher Eigen Vector Centrality scores.
This metric is calculated as the $v^{th}$ element of the vector $e$ derived from the equation:

    \begin{equation}
        Ae = \lambda e
    \end{equation}
Because all entries in matrix $A$ are non-negative, there is a distinct positive solution $e$ for the largest eigenvector $\lambda$.    

  \item Closeness Centrality: Closeness centrality identifies nodes that can efficiently disseminate information across a graph. This metric quantifies a node's average closeness (inverse of distance) to all other nodes. Nodes with high closeness scores are those with the shortest paths to other nodes. Closeness centrality for a node $u$ is the reciprocal of the average shortest distance to all other reachable nodes. It's calculated as:
    \begin{equation}
        c_u = \frac{n-1}{\sum_{l=1}^{n-1}d(u,l)}
    \end{equation}
    
here, $d(u, l)$ represents the shortest-path distance between nodes $u$ and $l$, and $n$ represents the number of nodes that can be reached from node $u$.
\item Clustering Coefficient:
The clustering coefficient of a node $u$ is a measure that reflects the proportion of triangles that involve the node. This can be expressed mathematically as:
 
    \begin{equation}
    \kappa_u = \frac{T(u)}{(deg(u))(deg(u)-1)}
    \end{equation}

In this equation, $T(u)$ represents the number of triangles passing through node $u$, and $deg(u)$ indicates the number of edges connected to that node.

\end{enumerate}

\vspace{2mm}
\subsubsection{Age group classification in the source domain}
Using the described feature extraction approach, each epoch yields $310$ features from $62$ scouts. The dataset includes $80$ epochs per subject, totaling $1200 $ epochs per stimulus type, equally distributed among Y, T, and M age groups. A $10$-fold cross-validation technique is employed to assess feature performance in age-group classification, reporting mean accuracy. Initially, the impact of varying the correlation threshold parameter ($\rho_{th}$)  on classification, utilizing the Random Forest (RF) classifier, is explored for different stimuli. Subsequently, the performance of alternative classifiers—Linear Support Vector Machines (Linear SVM), Logistic Regression (LR), and k-Nearest Neighbors (kNN)—is examined for varied stimuli. The classifiers use default parameters from the scikit-learn library \cite{sklearn_api}.

\subsection{Scout Connectivity Analysis}
The connectivity analyses involve extracting time series from source data (brain voxels or scouts), within a time window of -500ms to 900ms. These time series values were averaged for all the epochs in each stimulus case and are further used to compute an N*N correlation matrix per subject. The correlation matrix is computed after finding the mean of the time series. Then, the average connectivity matrices generated based on Pearson correlation for each age group (Y, T, and M) are compared. A connectivity graph is plotted using a chord diagram, where edge colors indicate Pearson correlation strength. Source-level connectivity was chosen due to the susceptibility of Sensor data to field spread and volume conduction across the scalp. Connectivity measures at the sensor level might misleadingly suggest brain connections. In contrast, connectivity measures between source time series are more anatomically interpretable and can be compared across participants. For frequency-specific brain connectivity analysis, the Phase Locking Value (PLV) \cite{tass1998detection} metric is used. PLV \cite{tass1998detection} leverages the relative instantaneous phase between two-time series to quantify connectivity. In source domain connectivity analysis, we study the functional connections between different brain lobes, which are categorized into 62 scouts, as illustrated in Figure 2.

\begin{figure*}[t]
\centering
\includegraphics[width =0.8\linewidth,keepaspectratio]{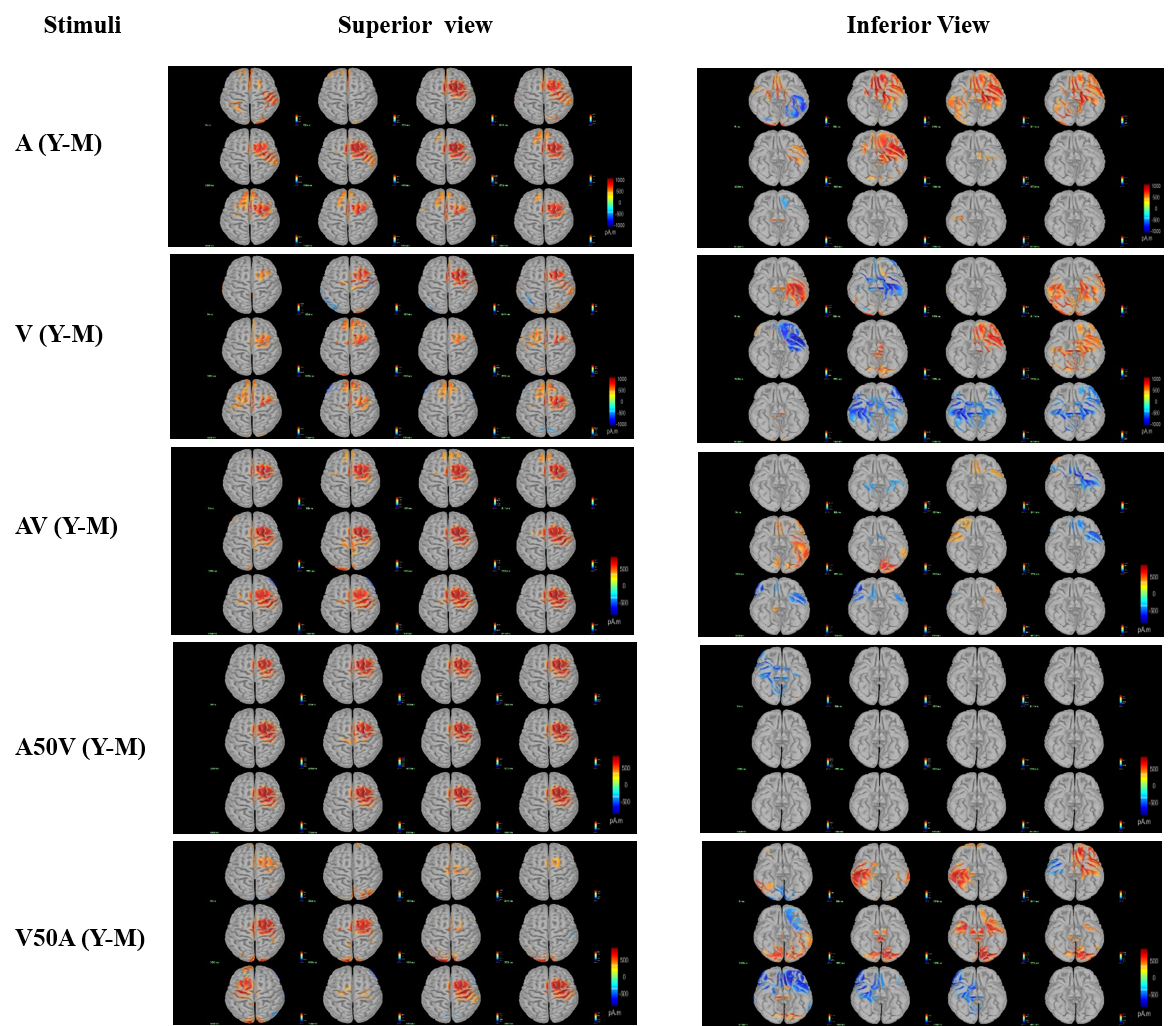}
\label{figure: figure 4}
\caption{Differences in scout activation and deactivation regions between young and middle-aged participants across varied stimulus conditions.}
\end{figure*}

\subsection{Scout Clustering}
In order to perform scout clustering for each stimuli condition, mean scout time series values for all the 62 scouts (Mindboggle atlas) were extracted within a time window of -500 ms to 900 ms. This generated a 62*700 dimensional matrix, which was further averaged for each and every subject. The resulting average matrix was used to calculate the average Pearson correlation matrix for every age group. Subsequently, it was transformed into Fisher’s z-transformed r-matrix using Fisher’s equation \cite{zar1996confidence}. The obtained r-matrix creates a comprehensive connected matrix, generating a graph of weighted relationships. Each participant's final data matrix was a 62*62 z-matrix with zero diagonal values. This matrix was subsequently used to cluster the 62 regions of interest (ROIs) into the optimal number of clusters. To achieve this, the K-means elbow algorithm \cite{marutho2018determination, singh2023reorganization} was applied to the average z-matrix of each age group, yielding an optimal configuration of five clusters. Using these clusters, nodes were rearranged to construct an adjacency matrix for each age group.

\begin{figure*}[t]
\centering
\includegraphics[width =0.5\linewidth,keepaspectratio]{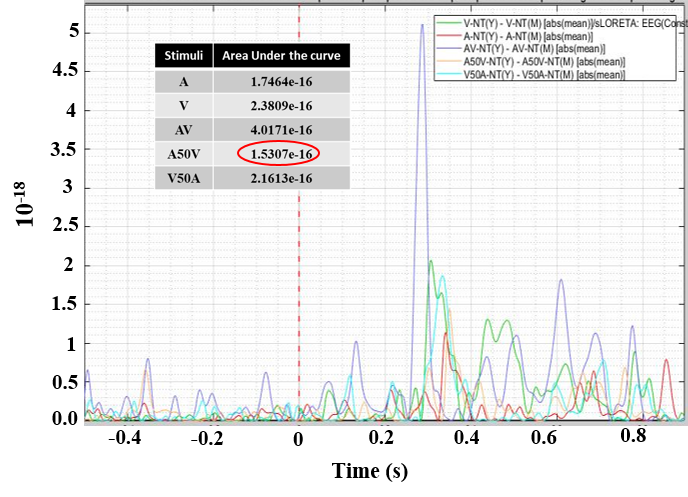}
\label{figure: figure 5}
\caption{Area under the peaks of AVI-associated regions for varied stimuli conditions.}
\end{figure*}

\section{Results}
\subsection{Impact of Ageing on the Brain Activation in Audio-Visual Integration?}
In the young age group, source activation is more pronounced in the early processing stages, particularly in frontal, temporal, and parietal brain regions. This early activation tends to diminish over time, especially in the theta frequency band. Activation is observed to start around 280-300 ms in response to stimuli and gradually decreases thereafter. This pattern is noticeable in both unimodal and bimodal stimuli cases. A comparison between unimodal and bimodal stimuli is illustrated in Figures 3a and 3b, respectively.
\par Whereas, in the middle age group, source activation begins slightly later, around 300-320 ms, and persists for a longer duration, particularly evident in bimodal cases such as simultaneous audio-visual (AV) and auditory lead visual by 50 ms (A50V) stimuli. Moreover, in bimodal cases, brain activation levels are generally higher in middle age compared to unimodal stimuli, suggesting an enhanced effect of audio-visual integration (AVI) with age.
 Also, there is an increased involvement of prefrontal, temporal, and parietal areas during AVI tasks in middle-aged participants. This indicates that middle-aged individuals exhibit higher brain activation in response to combined audio and visual stimuli, particularly in regions associated with integration, attention, and cognitive processing.

\subsection{Role of Lower Frequency Bands in Age-Related Brain Source Activation.}
 The study's findings reveal that in bimodal activation, the amplitude is higher in middle age compared to unimodal cases, as depicted in Figures 4a and 4b. These results specifically pertain to lower frequency bands, such as theta and alpha, as demonstrated in Figure 7. Higher frequency bands do not exhibit significant age-related changes. We performed a paired t-test to compare the unimodal and bimodal activation amplitude in the theta band for the Y, T, and M groups. The analysis revealed a statistically significant difference (t = -3.780, df = 2, p < 0.05 (p = 0.031), one-tailed), indicating that the mean of the bimodal (AV) group is significantly greater than that of the unimodal (V) group. This suggests that when both audio and visual stimuli are present, source activation levels increase with age. Notably, the theta band demonstrates higher source activation amplitude values than the alpha bands. The transition phase (25-33 years) shows the highest activation levels compared to the young and middle age groups in lower and higher frequency bands. In the theta bands, there is notably higher early activation, primarily in frontal, temporal, and parietal sites, which decreases with time. In the alpha bands, activation predominantly occurs in frontal and temporal sites, lasting longer in the middle age. 
\begin{figure*}[t]
\centering
\begin{subfigure}[h]{0.7\linewidth}
\includegraphics[width =\linewidth]{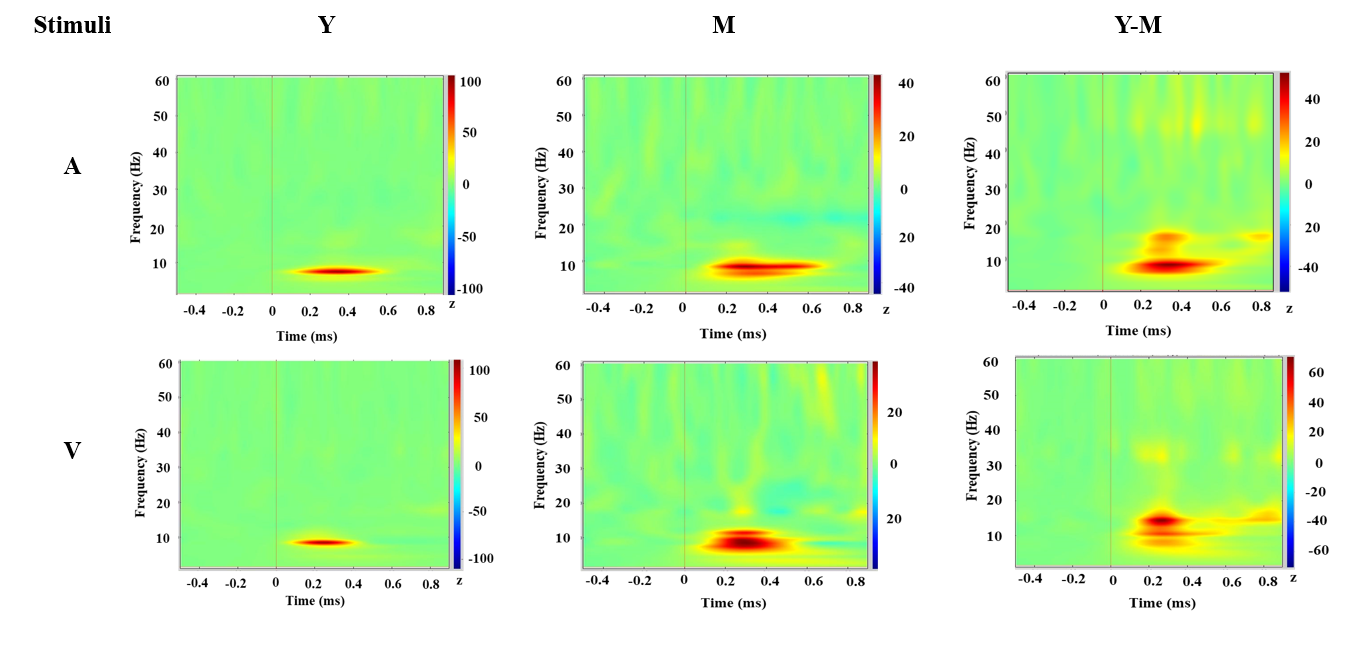}
\label{figure: figure 7a}
\caption{Time-Frequency decomposition in the case of unimodal stimulus.}
\end{subfigure}
\begin{subfigure}[h]{0.7\linewidth}
\includegraphics[width =\linewidth]{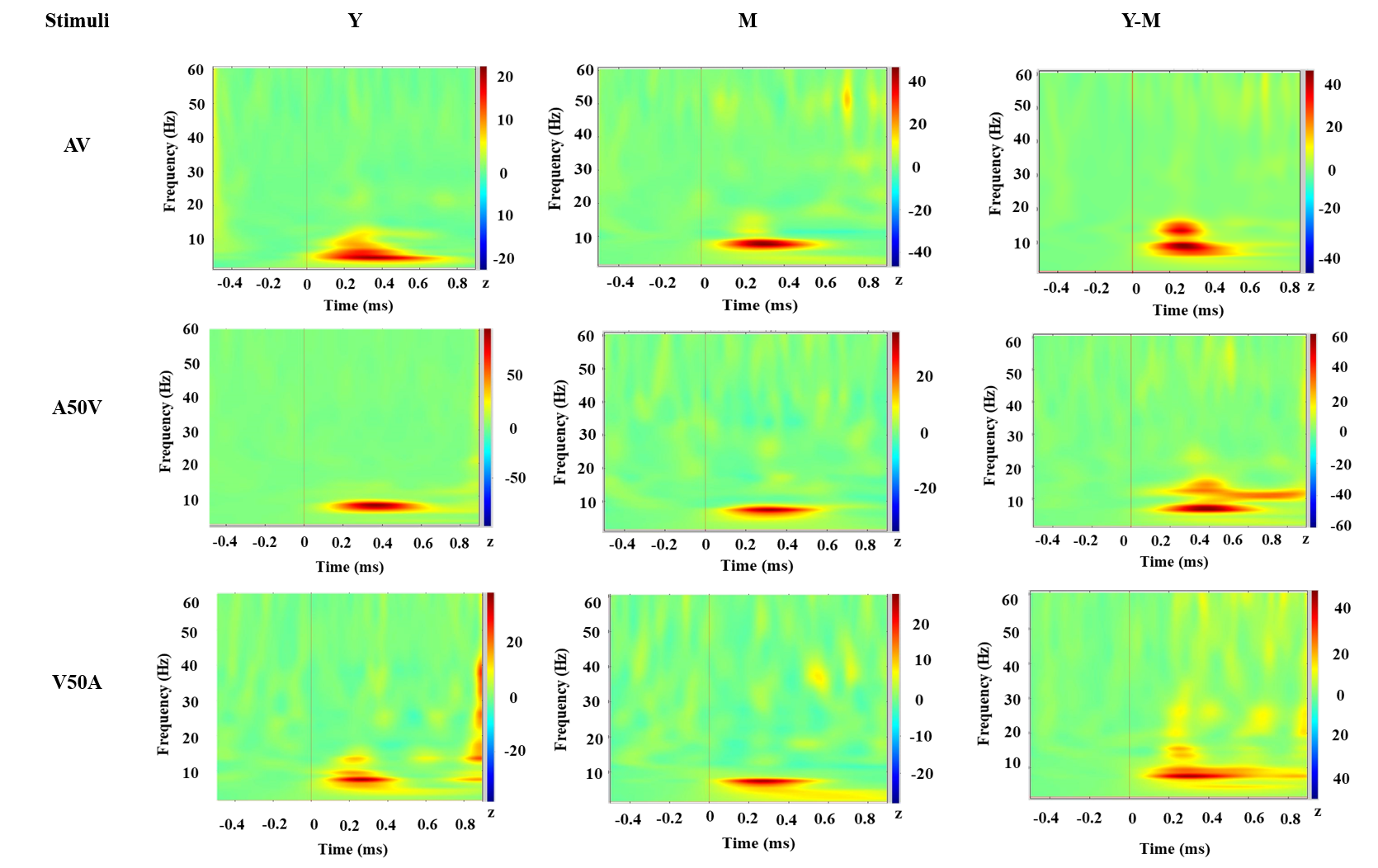}
\label{figure: Figure 7b}
\caption{Time-Frequency decomposition in the case of bimodal stimulus.} 
\end{subfigure}
\caption{Time-Frequency decomposition maps for young and middle-aged participants across varied stimulus conditions. The X-axis denotes the time axis and the Y-axis denotes the frequency bands.}
\end{figure*}
 
\subsection{Divergence in Brain Activation Sites in Response to Unimodal and Bimodal stimuli.}
Identification of significant AVI-related scouts was determined by examining the disparities in brain source activation due to aging. Scouts that exhibited noteworthy differences in activation levels across age groups were selected for further analysis. 
This selection process focused on pinpointing the scouts where the activation patterns underwent significant changes between the younger and middle-aged participants. 
In the case of unimodal stimuli, the difference between young and middle-aged brain source activation levels and the associated brain scouts is large as compared to bimodal stimuli as shown in Figure 5.
With the audiovisual integration effect, the differences in activation and deactivation levels decrease with aging. It reduces by $500 pAm$.
\par After analyzing the peaks of the Mindboggle scouts we observed that the significant scouts involved in the AVI effect are: caudal anterior cingulate, caudal middle frontal, fusiform, insula, lateral orbitofrontal, middle temporal R, parsopercularis, parstriangularis, superior frontal, superior parietal L, superior temporal and transverse temporal.
These scouts were merged together and labeled as AVI-associated scouts. Later the area under their peaks was calculated in the case of various stimuli for AVI-associated scouts. These peaks indicate the absolute mean difference in the average activation amplitude of young and middle-aged participants for the defined regions.
It was observed that the area under the curve in the case of A50V stimulus was the least. It indicates that the minimum difference between Y and M is in the case of A50V stimuli case which is clear from figure 5 and 6.

\subsection{Significant Differences in the Time-Frequency decomposition Maps with Aging.}
Here we can see the time-frequency decomposition of EEG data performed using Morlet wavelets. It is clear from Figures 7a and 7b, that there is a significant difference between young and middle-aged AVI-associated scout power maps in lower frequency bands. Scout mean power varies in both the unimodal and bimodal cases. Theta band shows major differences as the age increases followed by an alpha band. It concludes that theta and alpha bands are the most crucial frequency bands to study aging while performing Audio-visual tasks. Whereas higher frequency bands above 15 Hz are of typically very low power. Among bimodal stimuli, asynchronous stimuli show comparatively lower power difference values in lower frequency bands. It should be noted that the power scale varies in young and medium which resulted in differences in the power.

\subsection{Addition of Visual Stimuli to Auditory Stimuli Enhances Brain Functional Connectivity.}
Functional brain connectivity increases during middle age and these changes vary depending on the stimulus type. Most of the enhanced connections occur between brain regions in the prefrontal, frontal, temporal, limbic, and occipital areas. The introduction of visual stimuli significantly boosts brain connectivity. In particular, bimodal audio-visual stimuli show more extensive connectivity compared to unimodal auditory stimuli. Synchronous audio-visual stimuli exhibit more connections than asynchronous ones. Notably, the central and parietal lobes have fewer connections compared to the frontal lobes. Among the different stimulus types, visual stimuli (V) lead to the highest number of connections, followed by AV, V50A, A50V, and A. Therefore, incorporating visual stimuli into auditory stimuli enhances functional connectivity among various brain regions. Figure 8 shows increased brain functional connectivity in middle age during various stimuli at an optimum correlation threshold value $\rho_{th}=0.85$. This connectivity is plotted between different brain lobes subdivided into 62 Mindboggle scouts as depicted in Figure 2.

\begin{figure*}[t]
\centering
\includegraphics[width =0.8\linewidth,keepaspectratio]{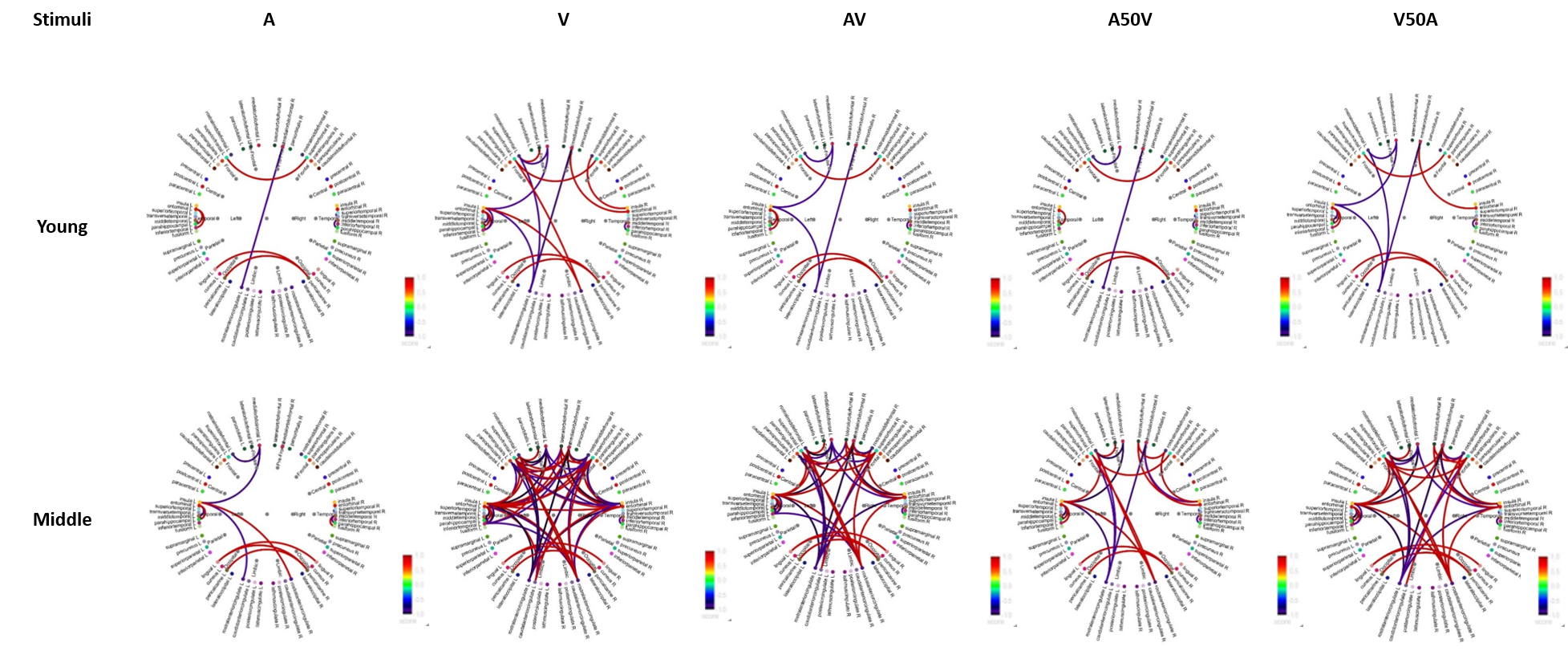}
\label{figure: figure 10}
\caption{Elevated brain functional connectivity in middle-aged individuals across different stimulus conditions at $\rho_{th}=0.85$.}
\end{figure*}

\begin{figure*}[t]
\centering
\begin{subfigure}[h]{0.8\linewidth}
\includegraphics[width =\linewidth]{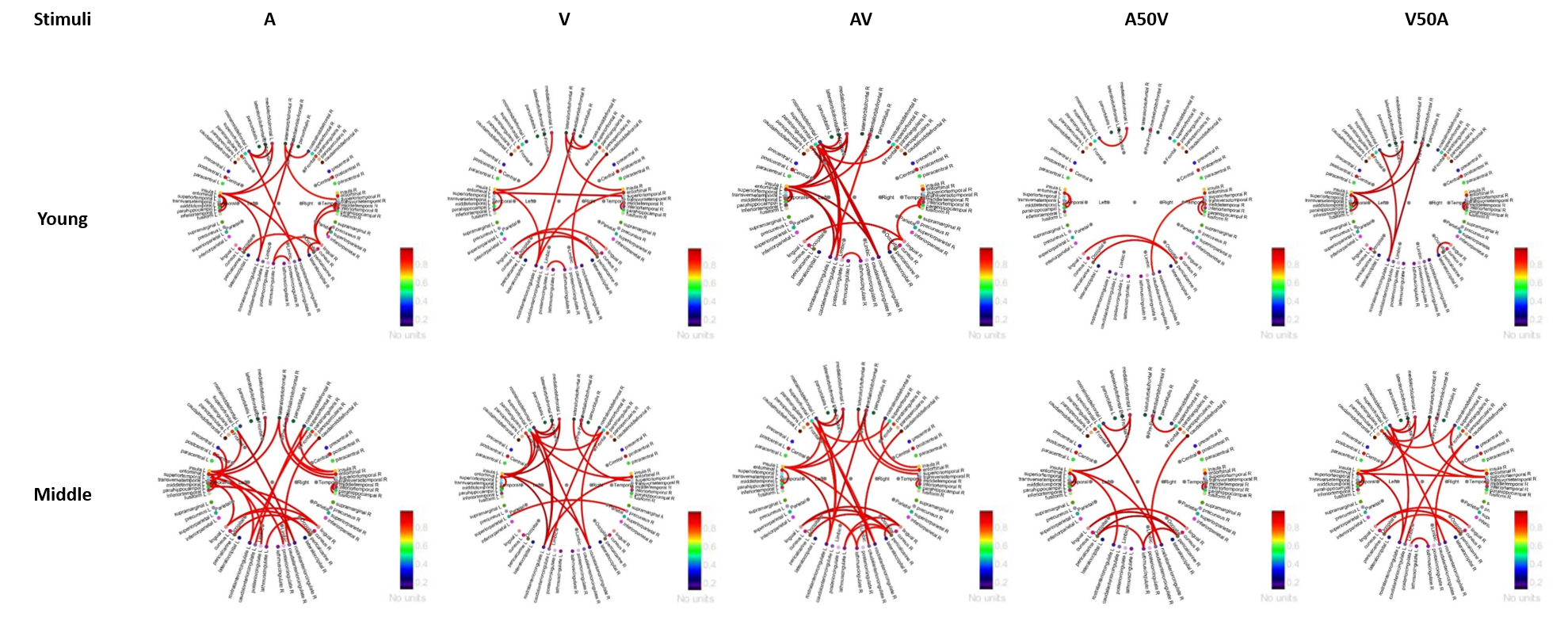}
\label{figure: figure 6a}
\caption{Brain functional connectivity in the theta band.}
\end{subfigure}
\begin{subfigure}[h]{0.8\linewidth}
\includegraphics[width =\linewidth]{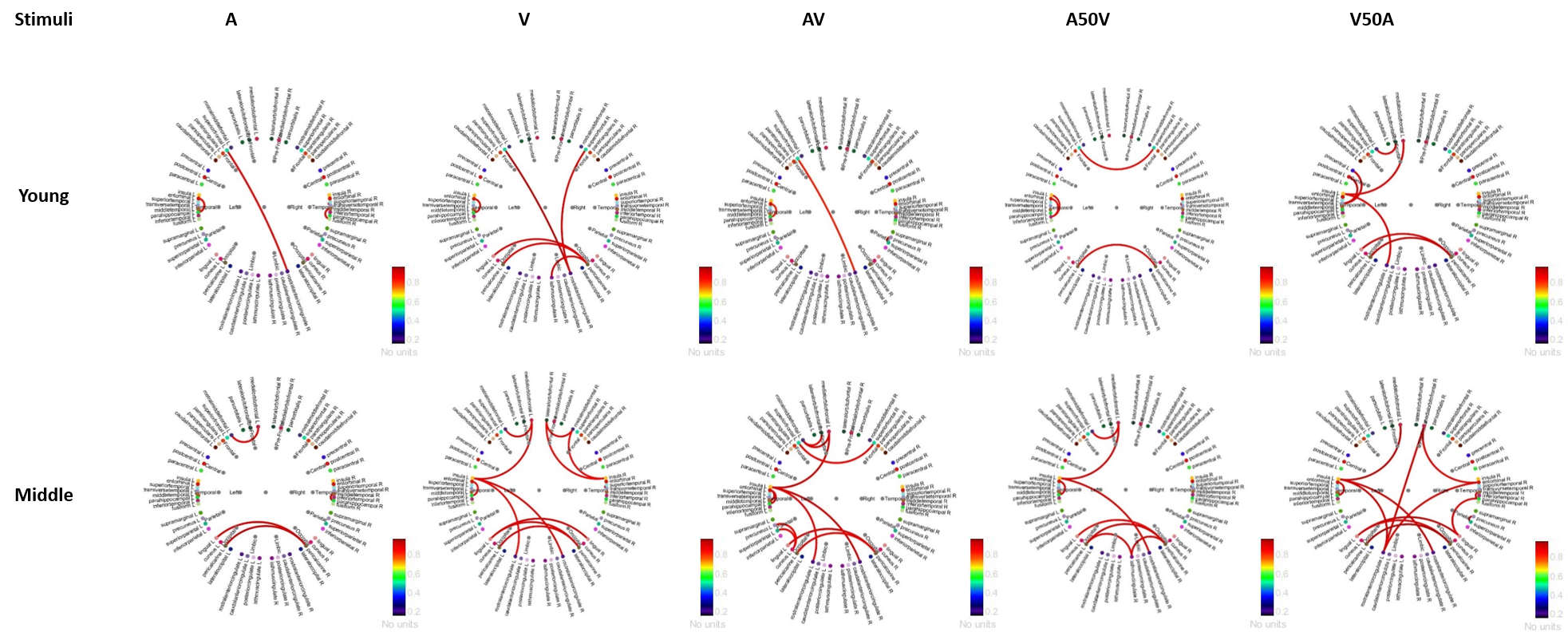}
\label{figure: Figure 6b}
\caption{Brain functional connectivity in the alpha band.} 
\end{subfigure}
\caption{Brain functional connectivity in theta and alpha bands for varied stimuli conditions.}
\end{figure*}

\subsection{ Theta Band exhibits the Highest Functional Connectivity During Audio-Visual Integration Tasks.}
In middle-aged individuals engaged in Audio-Visual Integration tasks, the theta band displays the highest functional connectivity. Higher frequency bands exhibit lower levels of brain source functional connectivity compared to lower frequency bands. These connections vary depending on the type of stimulus, as depicted in Figure 9a. Among all stimuli, synchronous AV stimuli feature the highest number of edges in the theta band, and this number increases with age, as illustrated in the figure.  The regions with the highest connectivity include the occipital, temporal, frontal, and pre-frontal areas. In the alpha band, visual (V), audio-visual (AV), and auditory lag visual by 50 ms (V50A) stimuli exhibit notable increases in connections with age, as indicated in Figure 9b.

\begin{figure*}[t]
\centering
\includegraphics[width =0.6\linewidth,keepaspectratio]{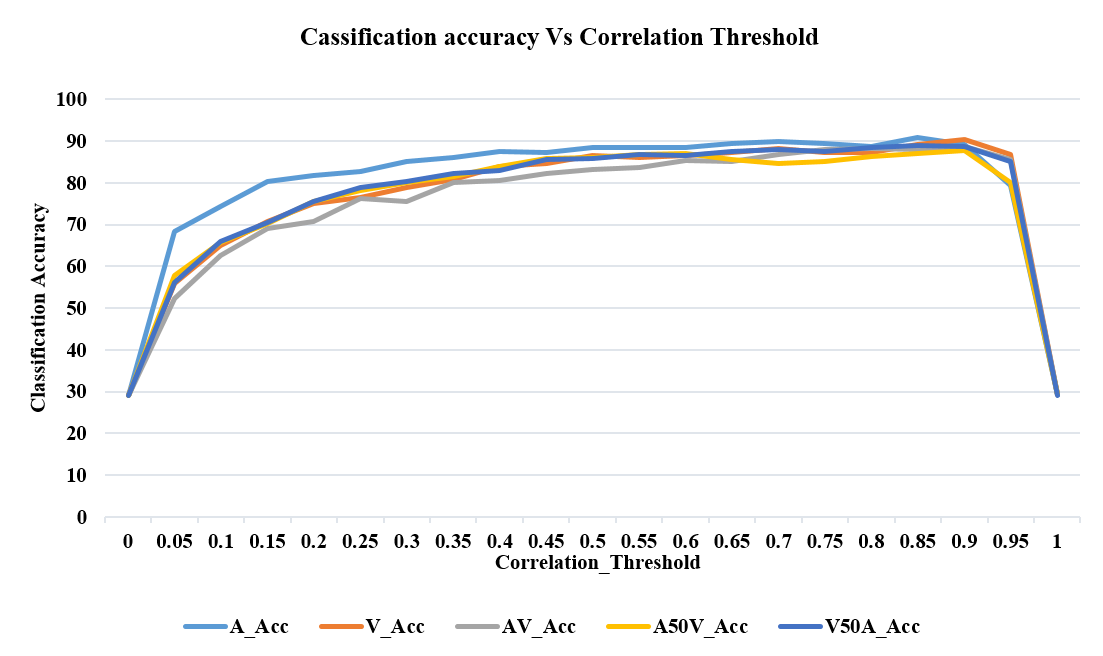}
\label{figure: figure 13}
\caption{Classification accuracy for different stimulus conditions at varying correlation threshold ($\rho_{th}$) value.}
\end{figure*}

\subsection{Scout Data-Based Age Group Classification.}
The impact of the correlation threshold ($\rho_{th}$) on classification accuracies for different stimulus types is illustrated in Figure 9 using a random forest classifier. Initially, the accuracy increases before decreasing. The optimal ($\rho_{th}$) value is found to be 0.85 and is used consistently thereafter for brain functional connectivity study. The Random Forest classifier achieves the highest classification accuracy of 90.75 \% for Audio stimuli, followed by 90.5 \% for V stimuli. This suggests that unimodal stimuli exhibit more pronounced differences compared to bimodal stimuli, resulting in lower classification accuracy for the latter. Among the bimodal stimuli, the V50A stimulus achieves the highest classification accuracy of 89\%, followed by AV (88.75 \%) and A50V (87.75 \%). After Random Forest, Linear SVM performs well followed by kNN and Linear Regression. Different classifiers were employed for age-group classification, and Table 1 provides an overview of the maximum accuracy achieved for each type of stimulus in the source domain. It is to be noted that the classification accuracies in the source domain for each stimulus are higher than the accuracies in the sensor domain \cite{singh2023brain}.

\begin{table}[!t]
\centering
\caption{Recognition accuracies for various classifiers.}
\centering
\scalebox{0.8}{
\begin{tabular}{ccccc}
\hline 
\textbf{Stimulus}  & \textbf{LinearSVM} & \textbf{kNN} & \textbf{LR} & \textbf{RF}                          \\\hline\hline
\textbf{A}       & 89.16              & 85.66        & {86.41}  & {\color[HTML]{0000FF}90.75}         \\
\textbf{V}       & 86.91              & 83.83        & {84.00}  & {\color[HTML]{0000FF}90.50}         \\
\textbf{AV}      & 87.08              & 83.58       & {84.00}    & {\color[HTML]{0000FF}88.75}       \\
\textbf{A50V}    & 85.75              & 80.08        & 83.30 & {\color[HTML]{0000FF}87.75}       \\
\textbf{V50A}    & 86.16             & 81.91        & {85.58}   & {\color[HTML]{0000FF}89.00}    \\\hline        
\end{tabular}
}
\label{tab:results}
\end{table}

\subsection{Clustering of Mindboggle Scouts within the Framework of AVI Task.}
The analysis results unveil five distinct clusters of brain networks derived from 62 brain scouts through k-means elbow clustering for each age group - young (Y), transition (T), and middle (M), as shown in Figure 11. These clusters, denoted as Cluster 1, Cluster 2, Cluster 3, Cluster 4, and Cluster 5, correspond to different brain regions. The clusters obtained can be shown in Figure 12.

\par A noteworthy observation is that brain networks or clusters exhibit heightened functional connectivity in middle-aged individuals compared to their younger counterparts. For instance, in the context of the AV stimulus, cluster 4 and cluster 1 demonstrate increased connectivity in middle age relative to young age, as depicted in Figure 13. Given that these clusters predominantly encompass frontal and temporal regions, it suggests that functional connectivity between the frontal and temporal lobes strengthens with age during tasks involving audio-visual integration. Consequently, brain network functional connectivity appears to undergo enhancement in middle age during audio-visual integration tasks.

\par Moreover, with respect to unimodal stimuli, brain functional connectivity reaches its peak in middle age during visual tasks. Concerning bimodal stimuli, the AV stimulus exhibits the highest functional connectivity, followed by the V50A stimulus, with age progression.

\par These findings provide insights into the dynamic alterations in brain connectivity across different stages of adulthood, offering valuable information regarding age-related distinctions in cognitive processing during audio-visual tasks.

\begin{figure*}[t]
\centering
\includegraphics[width =0.5\linewidth,keepaspectratio]{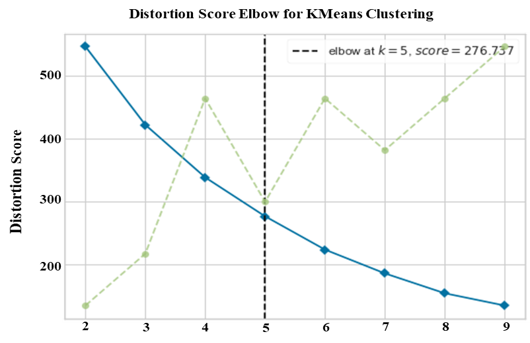}
\label{figure: figure 14}
\caption{Five optimum clusters obtained by clustering of 62 Mindboggle scouts.}
\end{figure*}

\begin{figure*}[t]
\centering
\includegraphics[width =0.98\linewidth,keepaspectratio]{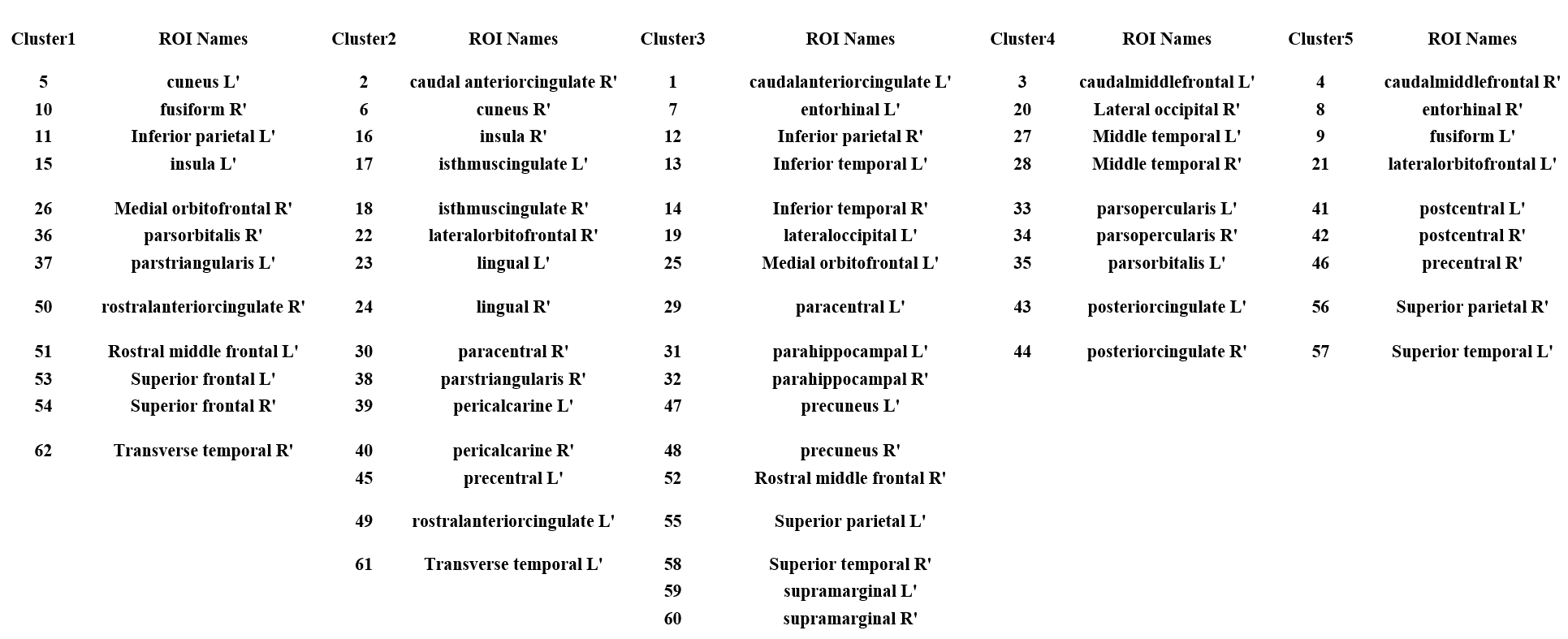}
\label{figure: figure 15}
\caption{Five brain networks obtained using a clustering approach.}
\end{figure*}

\section{Discussion}
  Neuroanatomical changes contribute to the cognitive declines during aging. The specific neuronal mechanisms involved in age-related audio-visual integration remain unclear. The present study has addressed this knowledge gap by investigating the effects of audio-visual integration on middle-aged individuals at the source level of brain activity. The research aimed to highlight the brain regions involved in this process and identify how their connectivity patterns evolve with age. The study also investigated the prominent EEG frequency bands linked to audio-visual tasks and their age-related dynamics. Additionally, it explored the dynamic brain networks formed during audio-visual integration and how their connectivity changes as individuals age.
 \par The study conducted an audio-visual integration (AVI) task to systematically examine middle-aged individuals' responses to audio and visual stimuli, including target and non-target stimuli. The findings revealed that middle-aged participants displayed increased brain activation when subjected to combined audio and visual stimuli. This heightened activation was particularly notable in brain regions associated with integration, attention, and cognitive processing. Important regions of interest included the caudal anterior cingulate, caudal middle frontal, fusiform, insula, lateral orbitofrontal, middle temporal (right), parsopercularis, parstriangularis, superior frontal, superior parietal (left), superior temporal, and transverse temporal areas. These regions were identified using the Brainstorm toolbox. These results align with prior research, including a meta-analytic study by Gao et al., conducted in 2022 \cite{gao2023audiovisual}. The outcomes of the present study underscore the pivotal role of the frontal, pre-frontal, temporal, and occipital lobes in the process of AVI during middle age, consistent with previous investigations \cite{shams2005early}.

 \begin{figure*}[t]
\centering
\includegraphics[width =0.8\linewidth,keepaspectratio]{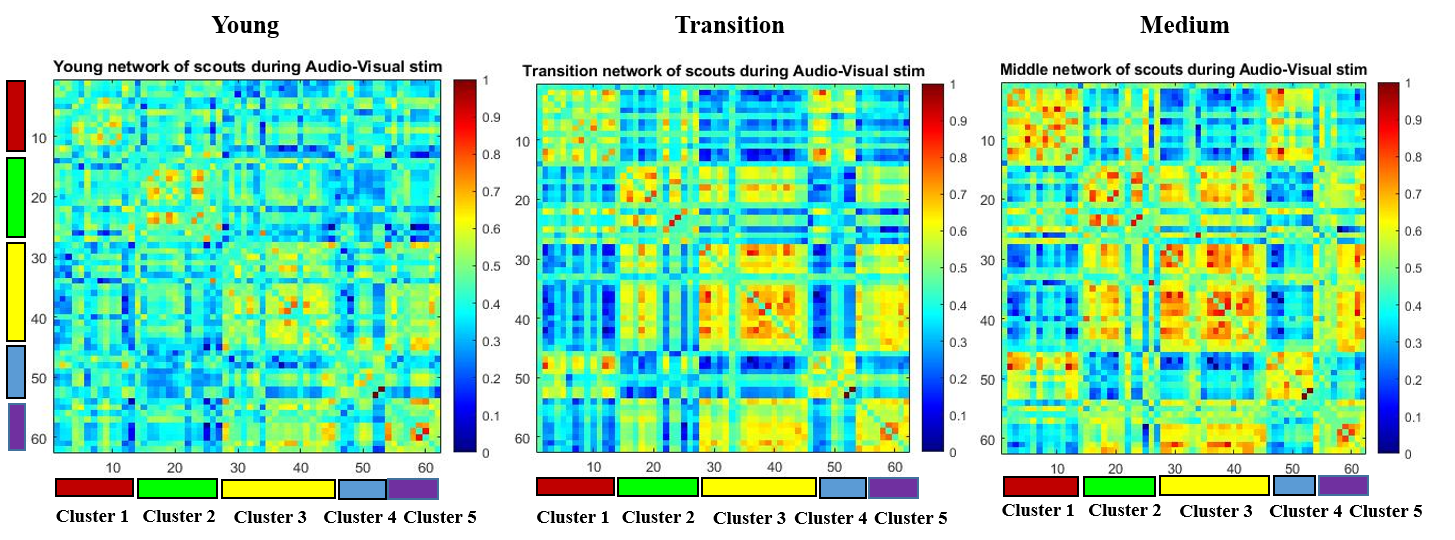}
\label{figure: figure 16}
\caption{Mean connectivity matrix, r(z) for different age cohorts based on 5 clusters during synchronous AV task.}
\end{figure*}

 \par 
 Our study discovered that middle-aged adults showed a weaker and delayed audio-visual integration (AVI) effect compared to younger adults in all conditions. This finding supports previous studies by Wu et al. (2012) and Ren et al. (2016), which observed similar trends \cite{ren2017audiovisual,wu2012age}. However, it's important to note that earlier research has produced conflicting results, with multiple reports presenting contradictory findings, as seen in studies like Laurienti et al. (2006) \cite{laurienti2006enhanced} and Diederich et al. (2008) \cite{diederich2008assessing}.
 It's important to note that some studies, that observed a stronger AVI effect in older adults, typically employed centrally presented stimuli. In contrast, our study used peripheral stimuli with a 5-degree visual angle. Peripheral vision tends to deteriorate with age \cite{anderson1997peripheral}, and this factor might play a role in the variations in our results.
 
\par Furthermore, with regard to the audio-visual integration (AVI) effect, our results highlight a decrease in the gap between brain activation and deactivation patterns in young and middle-aged adults. Notably, when considering unimodal stimuli, the difference in brain source activation levels between young and middle-aged individuals is more pronounced than with bimodal stimuli. This implies that the incorporation of visual stimuli alongside audio stimuli may partially alleviate the impact of aging on the brain. These findings strongly support the concept that audio-visual integration, as individuals age, functions as a compensatory mechanism, offsetting deficiencies in unimodal sensory processing \cite{davis2008pasa, mozolic2012multisensory}.
Our research highlights different patterns of activation in two frequency ranges: theta and alpha. In the theta range, we observe increased early activation, primarily in the frontal, temporal, and parietal areas, which decreases with age. In contrast, the alpha range shows continuous activation in the frontal and temporal regions, especially during middle age.
These findings align with prior EEG-based studies on aging, which have suggested diminished alpha power, indicating age-related variations in audio-visual integration, particularly in lower frequency bands \cite{kolev2002age}. Importantly, these patterns hold significance as both theta and alpha bands, particularly in frontal areas, are associated with cognitive control, short-term memory, and sensory information retention \cite{kawasaki2010dynamic}.
\par Our findings reveal higher brain functional connectivity in middle-aged and young adults under AV and V50A conditions compared to A and A50V conditions. This heightened connectivity is due to the synchronization of auditory and visual neural signals, which arrive at the brain with less temporal separation during AV and V50A conditions, resulting in a stronger integration effect. This aligns with previous research by \cite{molholm2002multisensory} which reported that visual stimuli have an onset latency of approximately 50 ms, while auditory stimuli have an onset latency of less than half that, around 9–15 ms from stimulus presentation. This closer temporal proximity likely amplifies the impact on the brain, leading to stronger connections between brain regions.
Moreover, we observed a significant increase in theta band functional connectivity in middle-aged adults during audio-visual stimuli. This increased connectivity predominantly occurs in pre-frontal, frontal, temporal, limbic, and occipital regions. Theta activity is associated with broader brain integration mechanisms and central executive functions during audio-visual integration \cite{sauseng2007dissociation, sauseng2010control,kawasaki2010dynamic}. Also, there is an increase in alpha band functional connectivity in middle age, particularly during AV and V50A conditions. The alpha activity reflects active attentional suppression mechanisms and executive functions \cite{kawasaki2010dynamic, kelly2006increases} both of which are vital for multisensory responses \cite{koelewijn2010attention}. Diederich et al. (2008) demonstrated that older adults exhibit greater neural enhancement for neural integration. This heightened cognitive demand with age likely contributes to the observed functional connectivity increases in the theta and alpha bands \cite{diederich2008assessing}. 
\par Our study has also revealed that brain networks formed during the audio-visual integration (AVI) effect can be categorized into five distinct networks. These networks exhibit variations in functional connectivity patterns with age. In summary, our findings align well with prior research, indicating that audio-visual integration experiences delays with age and displays differences between various age groups, such as young and old individuals \cite{ren2017audiovisual}.
Our results underscore the prominent involvement of the frontal and temporal cortex, particularly the superior temporal cortex, in AVI as individuals age. Furthermore, our study supports the notion of a compensatory neural mechanism with increasing age, characterized by heightened brain functional connections.
\par Nonetheless, our study has several limitations. Firstly, we utilized 32 scalp electrodes to construct brain networks in the source domain, which is a relatively small number. Future research could employ EEG devices with 64 or 128 channels to enhance the accuracy of our findings. Additionally, our study had a limited number of subjects in each age group, and expanding the sample size in future studies would be beneficial. We used a relatively small number of features for classification, which could be increased to improve accuracy. Expanding the age groups to include individuals in their late sixties would indeed contribute to a more comprehensive and comparative study. This additional age group can provide valuable insights into how audio-visual integration processes evolve in late adulthood, further enhancing our understanding of age-related cognitive changes in the source domain. Lastly, future research might consider conducting a longitudinal study on AVI in middle age to identify potential biomarkers that could aid in the early detection of brain abnormalities.

\section{Conclusion}
The study presents significant findings related to the audio-visual integration (AVI) process in different age groups under varied stimuli conditions. The differences in brain activation patterns, particularly in the superior frontal gyrus and superior temporal cortex, were observed between the young and middle-aged groups. In middle age, there was a delay in brain source activation, especially in response to bimodal stimuli.
The study also highlights the important role of lower frequency bands in both age groups during the Audio-visual integration. Furthermore, it validates an upsurge in functional connectivity within the brain during audio-visual integration, with these connections primarily concentrated in the frontal, temporal, limbic, and occipital areas.
These findings shed light on the compensatory neural mechanisms that come into play as individuals age, especially during specific cognitive tasks.

\section{Disclosure}
The authors have declared no conflict of interest related to this study.
  
\section{Acknowledgement}
  This research work is supported by the Neurocomputing Laboratory and Multichannel Signal Processing Laboratory (MSP Lab) at the Indian Institute of Technology Delhi (IIT Delhi), India. Data were collected at the Multichannel Signal Processing Laboratory (MSP Lab), IIT Delhi. The authors would like to thank all the participants for their contributions.

\bibliographystyle{IEEEtran}
\bibliography{refs.bib}

\end{document}